\documentstyle[graphics,graphicx,times]{aa}
\def\h2{H$_2$~}

\begin{document}

\title{Metals in the Intergalactic Medium\thanks{Based on observations
collected at the European  Southern Observatory (ESO), under the Large
Programme "The  Cosmic Evolution  of the IGM"  ID No.  166.A-0106 with
UVES on the 8.2m KUEYEN telescope operated at the Paranal Observatory,
Chile} }

\def\inst#1{{${}^{#1}$}}

\author{Bastien Aracil\inst{1}, Patrick Petitjean\inst{1,2}, 
Christophe Pichon\inst{3}, Jacqueline Bergeron\inst{1}\\
}
\offprints{B. Aracil}

\institute{$^1$ Institut d'Astrophysique de Paris -- CNRS, 98bis Boulevard 
Arago, F-75014 Paris, France, aracil@iap.fr\\
$^2$ LERMA, Observatoire de Paris-Meudon, 61 avenue de l'Observatoire, 
F-75014 Paris, France, petitjean@iap.fr\\
$^3$ Observatoire de Strasbourg, 11 rue de l'Universit\'e,
67000 Strasbourg, France. \\
}

\date{Typeset \today ; Received / Accepted}

\abstract{  We use  high  spectral resolution  ($R$~=~45000) and  high
signal-to-noise  ratio  (S/N~$\sim$~35-70  per  pixel) spectra  of  19
high-redshift  (2.1~$<$~$z_{\rm em}$~$<$~3.2)  quasars  to investigate
the  metal  content  of  the low-density  intergalactic  medium  using
pixel-by-pixel procedures.  This high quality homogeneous survey gives
the  possibility  to statistically  search  for  metals  at H~{\sc  i}
optical depths smaller  than unity.  We find that  the gas is enriched
in carbon  and oxygen for  neutral hydrogen optical  depths $\tau_{\rm
HI}$~$>$~1.   Our  observations   strongly  suggest  that  the  C~{\sc
iv}/H~{\sc  i} ratio  decreases with  decreasing $\tau_{\rm  HI}$ with
$\log\,\tau_{\rm  CIV}\,=\,1.3\times\log\,\tau_{\rm  HI}-3.2$.  We  do
not detect C~{\sc iv}  absorption statistically associated with gas of
$\tau_{\rm HI}$~$<$~1.   However, we observe that a  small fraction of
the  low  density gas  is  associated with  strong  metal  lines as  a
probable   consequence   of    the   IGM   enrichment   being   highly
inhomogeneous. We detect the presence of O~{\sc vi} down to $\tau_{\rm
HI}$~$\sim$~0.2       with       log~$\tau_{\rm       OVI}$/$\tau_{\rm
HI}$~$\sim$~$-$2.0.  We show that  O~{\sc vi} absorption in the lowest
density gas is located within $\sim$300~km~s$^{-1}$ from strong H~{\sc
i} lines.  This suggests  that this  O~{\sc vi} phase  may be  part of
winds  flowing  away from  overdense  regions.   This  effect is  more
important at  the largest redshifts ($z$~$>$~2.4).   Therefore, at the
limit of  present surveys,  the presence of  metals in  the underdense
regions of the IGM is still to be demonstrated.
\keywords{
{\em Cosmology:} observations -- {\em Galaxies:} haloes 
-- {\em Galaxies:} ISM
-- {\em Quasars:} absorption lines
-- {\em Quasars:} individual: 
}}

\maketitle

\section {Introduction}
\label{s:introduction}
One of the key issues  in observational cosmology is to understand how
and when star formation took  place in the high redshift universe.  In
particular, it is not known when  the first stars appeared or how they
were  spatially distributed. The  direct detection  of these  stars is
challenging  but the intergalactic  medium (IGM)  provides at  least a
record of stellar  activity at these remote times.  Indeed, metals are
produced in stars  and expelled into the IGM  by supernovae explosions
and subsequent  winds and/or by  galaxy interactions. It  is therefore
crucial to  observe the distribution of  metals present in  the IGM at
high redshifts.
\par\noindent
The high-redshift  intergalactic medium (IGM) is  revealed by numerous
H~{\sc i} absorption  lines observed in the spectra  of remote quasars
(the so-called Lyman-$\alpha$ forest). It  is believed that the gas in
the IGM traces the potential wells  of the dark matter and its spatial
structures:  overdense  sheets   or  filaments  and  underdense  voids
(e.g. Cen et  al. 1994, Petitjean et al. 1995,  Hernquist et al. 1996,
Bi \& Davidsen  1997).  In the course of cosmic  evolution, the gas is
most  likely metal  enriched by  winds flowing  out  from star-forming
regions  that are  located  preferentially in  the  centre of  massive
halos.   It  is  therefore   not  surprising  to  observe  C~{\sc  iv}
absorption associated  with most  of the strong  H~{\sc i}  lines with
log~$N$(H~{\sc i})~$>$~14.5 as these lines most likely trace filaments
in  which massive halos  are embedded  (Cowie et  al. 1995,  Tytler et
al. 1995).   The question  of whether the  gas filling  the underdense
space (the  so-called voids) delineated by  these overdense structures
also contains metals or not  is crucial. Indeed, it is improbable that
winds from  star-forming regions located in the  filaments can pollute
the voids  entirely (Ferrara  et al. 2000).  Therefore, if  metals are
found in the gas filling the  voids, then they must have been produced
in the very early Universe by objects more of less uniformly spatially
distributed.\par\noindent
\par\noindent
Absorptions arising  through voids are mostly  of low-column densities
(typically    of     the    order    or     less    than    $N$(H~{\sc
i})~=~10$^{13}$~cm$^{-2}$).   Given   the  expected  low  metalicities
(typically  [C/H]~$<$~$-$2.5 relative to  solar), direct  detection of
metals  at  such  low  neutral  hydrogen optical  depth  is  currently
impossible due  to the weakness  of the expected metal  absorption and
statistical methods should be used  instead. Lu et al. (1998) used the
stacking  method to increase  the signal-to-noise  ratio at  the place
where metal absorptions are expected and did not find any evidence for
metals        in        the        range        10$^{13}$~$<$~$N$({\rm
HI)~$<$~10$^{14}$~cm$^{-2}$.   Although uncertainties in  the position
of the lines  can lead to underestimate the  absorption, they conclude
that metalicity is smaller than 10$^{-3}$~solar in this gas. Note that
this limit has been confirmed by  Ellison et al. (2000) using the same
method. Cowie  \& Songaila (1998) introduced  another method measuring
the mean C~{\sc  iv} optical depth corresponding to  all pixels of the
Lyman-$\alpha$  forest  with  similar  H~{\sc i}  optical  depth  (see
Aguirre et al.  2002 for  an extensive discussion of the method). They
showed  that  the  mean  C~{\sc  iv}  optical  depth  correlates  with
$\tau_{\rm HI}$ for $\tau_{\rm HI}$~$>$~1.  Ellison et al. (2000) used
the same method  on a spectrum of very high signal  to noise ratio and
concluded  that  the  data  are  consistent with  an  almost  constant
log~C~{\sc    iv}/H~{\sc    i}~$\sim$~$-$3    down    to    $\tau_{\rm
HI}$~$\sim$~2$-$3.   Using  a large  sample  of  high resolution  data
Schaye  et al.  (2003) find  that  the carbon  abundance is  spatially
highly inhomogeneous. From simulations including strong assumptions on
the  UV  background  they  conclude  that the  median  metallicity  is
[C/H]~=~$-$3.47$^{+0.08}_{-0.06}$.
\par\noindent
The method was applied to search for O~{\sc vi} by Songaila (1998) who
found a  significant amount  of O~{\sc  vi} in gas  with a  C~{\sc iv}
optical  depth greater than  0.05, and  by Schaye  et al.   (2000) who
claimed to have  detected O~{\sc vi} in gas of  mean H~{\sc i} optical
depth as low  as 0.1.  However, Petitjean (2001) has  shown that a non
negligible  fraction of  the signal  could originate  in the  wings of
strong  lines  which  are   associated  with  overdense  regions.   In
addition,  note  that  three  different groups  have  found  different
results concerning the nature of the O~{\sc vi} phase (Carswell et al.
2002, Simcoe et al. 2002, Bergeron et al. 2002).
\par\noindent

We have applied the methods introduced by Cowie \& Songaila (1998), in
a slightly  modified form, to a  set of homogeneous data  of very high
quality obtained in the course  of the ESO Large Programme "The Cosmic
Evolution of  the IGM",  in order  to search for  both C~{\sc  iv} and
O~{\sc  vi}  absorptions  in  the IGM  at  $z$~$\sim$~2.5.   Section~2
describes  the   data  and  Section~3  the  method,   results  of  the
investigation are presented in Section~4.
\section{Data}
The ESO-VLT Large Programme "The Cosmic Evolution of the IGM" has been
devised to gather an homogeneous sample of lines of sight suitable for
studying the  Lyman-$\alpha$ forest  in the redshift  range 1.7$-$4.5.
High  spectral  resolution  ($R$~$\sim$~45000),  high  signal-to-noise
ratio ($\sim$35 and 70 per  pixel at, respectively, 3500 and 6000~\AA)
UVES spectra have been taken over the wavelength ranges 3100--5400 and
5450--9000~\AA.   Although  the   complete  emission  redshift  range,
2.2$-$4.5, is covered,  emphasise is given to lower  redshifts to take
advantage of the very good sensitivity  of UVES in the blue and of the
fact  that the  Lyman-$\alpha$ forest  is less  blended  and therefore
easier to  analyse.  In particular,  metal lines and amongst  them the
important O~{\sc vi} transitions can be more easily detected.
\par\noindent
Observations have  been performed  in service mode  over 4  periods (2
years). Details of data reduction and procedures used to normalise the
spectra  and preanalyse  metal lines  automatically will  be described
elsewhere. In  brief, the data are  reduced using the  UVES context of
the ESO  MIDAS data reduction package applying  the optimal extraction
method  and  following  the  pipeline  reduction step  by  step.   The
extraction  slit length  is  adjusted to  optimise the  sky-background
subtraction.    The   procedure   systematically  underestimates   the
sky-background  signal but  the  final accuracy  is  better than  1\%.
Addition of  individual exposures is performed using  a sliding window
and weighting the signal by the total errors in each pixel.
An  automatic   procedure  estimates  iteratively   the  continuum  by
minimising the sum of a regularisation term and a $\chi^2$ term, which
is computed  from the difference  between the quasar spectrum  and the
continuum estimated  during the previous  iteration.  Absorption lines
are avoided using the estimated  continuum.  A few obvious defects are
then corrected by hand adjusting the reference points of the fit. This
happens  to  be  important   in  small  regions  common  to  different
observational settings and near the peak of strong emission lines. The
automatic method works very well however and manual intervention is at
the minimum.
We  have carefully  calibrated this  procedure using  simulated quasar
spectra  (with   emission  and  absorption   lines)  adding  continuum
modulations to  mimic an imperfect  correction of the blaze  along the
orders and noise to obtain a S/N ratio similar to that in the data. We
noted  that the  procedure underestimates  the true  continuum  in the
Lyman-$\alpha$  forest  by  a   quantity  depending  smoothly  on  the
wavelength and  the emission  redshift and amounting  to about  2\% at
$z$~$\sim$~2.3.   To calibrate  this  quantity we  have simulated  QSO
absorption spectra drawing H~{\sc i} absorption lines at random from a
population  with  the  same   column  density  and  Doppler  parameter
distributions as observed, using  the spectral resolution and noise as
in  the UVES  data.  We  apply  the same  normalisation procedure  and
compute the ratio  of the input to the output  continua. This ratio is
plotted in Fig.~\ref{corrlya} for different emission redshifts.
\begin{figure}
\resizebox{0.9\hsize}{!}{\includegraphics{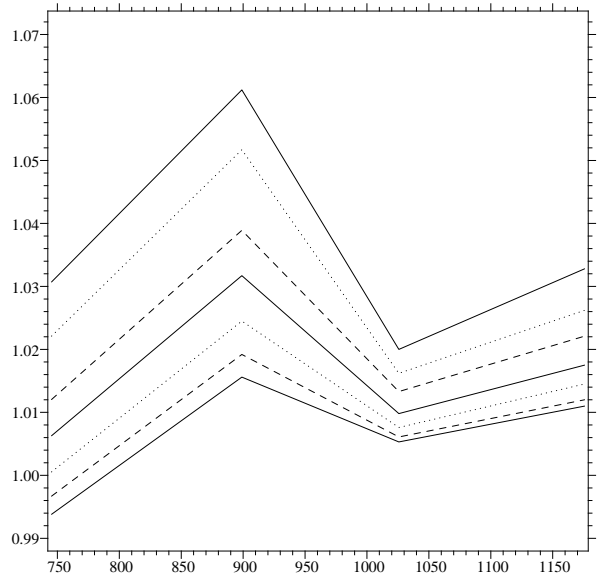}}
\caption{Ratio  of the  true continuum  to the  recovered one  used to
correct  the  estimated  continuum  as  a function  of  wavelength  as
measured  in the  rest-frame  of  the quasar.   This  is derived  from
normalization  of simulated  spectra. Different  curves  correspond to
different  redshifts ($z$~=~2.0,  2.1, 2.3,  2.5, 2.7,  3.0,  3.2 from
bottom to top).}
\label{corrlya}
\end{figure}

\par\noindent
Here we use nineteen lines of sight that are described in Table~\ref{objects}
and Fig.~\ref{limits}. The mean redshift of the survey is approximately
$z$~$\sim$~2.4.
\begin{table}
\caption{List of lines of sight \label{objects}}
\begin{scriptsize}
\begin{minipage}{\hsize}
\renewcommand{\footnoterule}{}
\begin{tabular*}{1.0\hsize}{
    l@{\extracolsep{\fill}} 
    l@{\extracolsep{\fill}}
    c@{\extracolsep{\fill}}
    c@{\extracolsep{\fill}}
    c@{\extracolsep{\fill}}r} \hline\hline
  Name & $z_{\rm em}$  & & Coverage\footnote{\scriptsize The upper limit
    is the  emission redshift blue-shifted by  3000~km~s$^{-1}$.  The lower
    limits for the different redshift coverages (respectively, 
    the forest, C~{\sc  iv} and O~{\sc vi} coverages) 
    are the minimal  observable redshift  (due to  instrumental limitation  or 
    the presence of a Lyman
    limit  break) for  Lyman$\alpha$, Lyman$\beta$  and  O~{\sc vi}
    respectively. Lyman$\beta$ is used  for the C~{\sc iv} coverage to
    be sure that the H~{\sc i} optical depth associated  to C~{\sc iv} is
    derived from at least two Lyman series lines. The SNR per pixel is averaged over 
    the C~{\sc iv} regions.} & \\
    &            &Forest&C~{\sc iv}&O~{\sc vi}&SNR\\
\hline
HE1341$-$1020    &  2.135  &  1.55$-$2.10 &   2.02$-$2.10 &   2.00$-$2.10 &  66\\
Q0122$-$380      &  2.190  &  1.59$-$2.16 &   2.07$-$2.16 &   2.05$-$2.16 &  64\\
PKS1448$-$232    &  2.220  &  1.59$-$2.19 &   2.07$-$2.19 &   2.05$-$2.19 &  69\\
PKS0237$-$23     &  2.222  &  1.55$-$2.19 &   2.02$-$2.19 &   2.00$-$2.19 & 117\\
HE0001$-$2340    &  2.263  &  1.55$-$2.23 &   2.02$-$2.23 &   2.00$-$2.23 &  98\\
Q0109$-$3518     &  2.404  &  1.59$-$2.37 &   2.07$-$2.37 &   2.05$-$2.37 & 105\\
Q1122            &  2.410  &  1.55$-$2.38 &   2.02$-$2.38 &   2.00$-$2.38 &  63\\
HE2217$-$2818    &  2.414  &  1.55$-$2.38 &   2.02$-$2.38 &   2.00$-$2.38 &  88\\
Q0329$-$385      &  2.435  &  1.55$-$2.40 &   2.02$-$2.40 &   2.00$-$2.40 &  61\\
HE1158$-$1843    &  2.449  &  1.55$-$2.41 &   2.02$-$2.41 &   2.00$-$2.41 &  76\\
HE1347$-$2457    &  2.611  &  1.55$-$2.58 &   2.02$-$2.58 &   2.00$-$2.58 & 104\\
Q0453$-$423      &  2.658  &  1.59$-$2.62 &   2.07$-$2.62 &   2.05$-$2.62 &  84\\
PKS0329$-$255    &  2.703  &  1.62$-$2.67 &   2.11$-$2.67 &   2.09$-$2.67 &  45\\
Q0002$-$422      &  2.767  &  1.63$-$2.73 &   2.12$-$2.73 &   2.10$-$2.73 &  69\\
HE0151$-$4326    &  2.789  &  1.63$-$2.75 &   2.12$-$2.75 &   2.11$-$2.75 &  97\\
HE2347$-$4342    &  2.871  &  1.87$-$2.83 &   2.40$-$2.83 &   2.38$-$2.83 &  75\\
HE0940$-$1050    &  3.084  &  1.96$-$3.04 &   2.51$-$3.04 &   2.49$-$3.04 &  84\\
Q0420$-$388      &  3.117  &  2.09$-$3.08 &   2.67$-$3.08 &   2.64$-$3.08 &  88\\
PKS2126$-$158    &  3.280  &  2.04$-$3.24 &   2.60$-$3.24 &   2.58$-$3.24 &  79\\
\hline
\end{tabular*}
\end{minipage}
\end{scriptsize}\label{tab:1}
\end{table}
\begin{figure}
\centering
\resizebox{0.9\hsize}{!}{\includegraphics{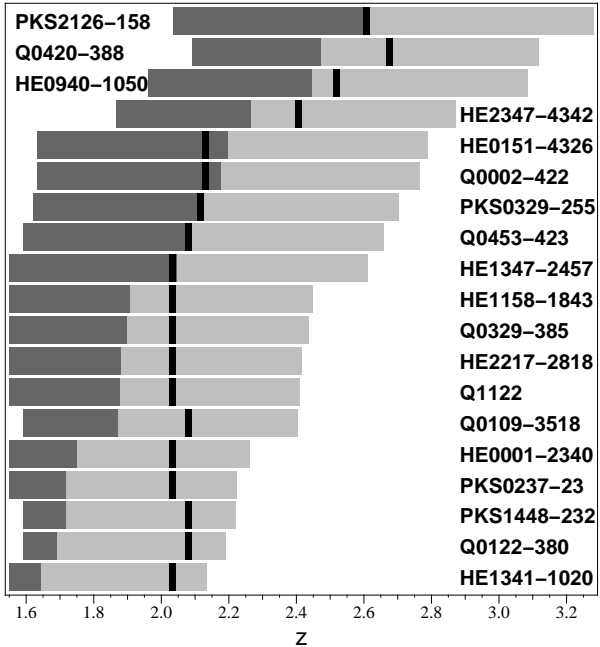}}
\caption{Limits of the Lyman-$\alpha$  forest along each line of sight
in  the  survey. Different  colors  correspond  to different  redshift
ranges.    Light   grey  corresponds   to   the   range  between   the
Lyman-$\alpha$   and   Lyman-$\beta$   emission   lines   where   only
Lyman-$\alpha$ absorption  is expected.  Dark grey corresponds  to the
range where Lyman-$\alpha$ is blended  with other lines in the series.
Black ticks  correspond to the  redshift at which Lyman$\beta$  is not
redshifted enough to be observed.}
  
\label{limits}
\end{figure}
\section{Pixel optical depth method}
In the highest  SNR QSO spectrum obtained up to  now, metals have been
identified by direct detection of  C~{\sc iv} absorption lines down to
column densities  as small as  $N$(C~{\sc iv})~=~10$^{11.7}$~cm$^{-2}$
(Ellison et al. 2000).  Most of these lines are associated with H~{\sc
i} column densities larger  than 10$^{14}$~cm$^{-2}$ and are therefore
believed  to trace  overdense regions  such  as filaments  of the  IGM
web.   Our   data   have   usually   a  detection   limit   of   about
10$^{12}$~cm$^{-2}$   in   the   C~{\sc   iv}   region   outside   the
Lyman-$\alpha$ forest.  To test for the  presence of metals  in gas of
smaller H~{\sc i} optical depth, we use a variant of the pixel optical
depth method (POD),  that associates the H~{\sc i}  optical depth with
that in metals (mainly C~{\sc iv}  and O~{\sc vi}) on a pixel by pixel
basis  (Cowie  \& Songaila  1998,  Ellison  et  al. 2000,  Aguirre  et
al. 2002).
\par\noindent
The principle  of the method  is as follows.  Once  the Lyman-$\alpha$
forest  is  cleaned  from  all  metals  (see  below),  the  H~{\sc  i}
Lyman-$\alpha$  optical  depth is  calculated  in  each  pixel of  the
forest. When the Lyman-$\alpha$ line is saturated, other transition in
the  Lyman  series  are  used.   For each  Lyman-$\alpha$  pixel,  the
observed optical depth in the metal transition at the same redshift is
measured.  Pixels in the Lyman-$\alpha$ forest are sorted according to
their Lyman-$\alpha$  optical depth and are gathered  in optical depth
bins that contains  a large number of pixels.  Then  the median of the
associated  metal  optical  depths  is calculated  in  the  predefined
bins. The median  is chosen instead of the mean to  avoid a few pixels
with high optical  depth to bias the mean optical  depth.  In that way
it  is possible  to test  for the  overall presence  of metals  in low
density  gas. The  method has  been extensively  tested by  Aguirre et
al.  (2002).  Here  we improve  the method  by carefully  cleaning the
Lyman-$\alpha$ forest for the presence of metals.

\par\noindent
\begin{figure}
\centering
\resizebox{0.9\hsize}{!}{\includegraphics{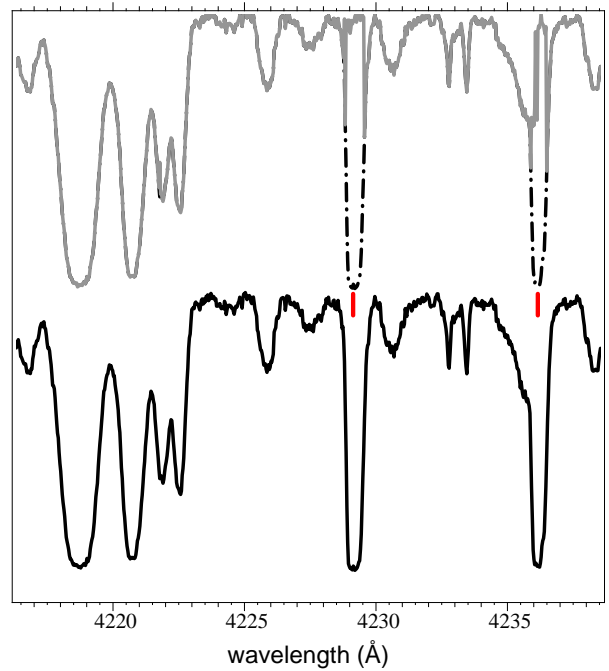}}
\caption{The  Lyman-$\alpha$ forest  (lower  curve; here  part of  the
HE~0151$-$4326  spectrum)  is  fitted  consistently  using  the  Lyman
series.  The  automatic procedure  is  therefore  able  to detect  the
presence of metals.  The upper curve shows the fit of the forest where
the procedure  has recognised  the presence of  the strong  C~{\sc iv}
system indicated by tick marks.  }
\label{filter}
\end{figure}
\subsection{Determination of $\tau_{\rm HI}$ and cleaning the 
Lyman-$\alpha$ forest from metals}
We want  to estimate the  Lyman-$\alpha$ optical depth in  the largest
number of pixels possible. For this  we have to take into account that
the absorption optical depth in any pixel can be due to several H~{\sc
i} Lyman transitions at different  redshift and also to possible metal
lines associated with strong systems.  If $z_{\rm em}$ is the emission
redshift  of  the  quasar,   and  1215.67  and  1025.67~\AA~  are  the
wavelengths  of   the  H~{\sc  i}   Lyman-$\alpha$  and  Lyman-$\beta$
transitions,     in     the     wavelength    range     [$\lambda_{\rm
min}$,$\lambda_{\rm                max}$]~=~               [(1+$z_{\rm
em}$)$\times$1025.67,(1+$z_{\rm       em}$)$\times$1215.67],      only
Lyman-$\alpha$  absorption  is  expected   from  H~{\sc  i}.   In  the
wavelength  range  $\lambda$~$<$~(1+$z_{\rm  em}$)$\times$1025.67  the
total  absorption is due  to the  Lyman-$\alpha$ forest,  the possible
metal lines  and also  the absorptions in  the different  Lyman series
transitions from the Lyman-$\alpha$ forest at higher redshift.
\par\noindent
We start considering $\lambda^{\rm obs}_{\rm alpha}$ in the wavelength
range   [$\lambda_{\rm   min}$,$\lambda_{\rm   max}$]~=~   [(1+$z_{\rm
em}$)$\times$1025.67,(1+$z_{\rm   em}$)$\times$1215.67].    For   each
pixel,       corresponding       to      $\lambda^{\rm       obs}_{\rm
alpha}$~=~(1~+~$z$)$\times$1215.67,  we measure the  effective optical
depth  in  Lyman-$\alpha$ and  calculate  the corresponding  effective
optical   depths  at  the   positions,  (1~+~$z$)$\times$$\lambda^{\rm
0}_{\rm series}$, of all transitions in the Lyman series redshifted in
the observed wavelength range,
\begin{equation}
\tau_{\rm series} = - {f_{\rm series}\over f_{\rm alpha}}
ln[F( {\lambda^{\rm obs}_{\rm alpha}\over 1215.67}\lambda^{\rm 0}_{\rm series})]
\end{equation}
where  $f_{\rm  series}$ is  the  transition  oscillator strength  and
$\lambda^{\rm  0}_{\rm  series}$  the  laboratory  wavelength  of  the
transition in the series. When the flux in a pixel is smaller than the
noise rms  at this position, we  flag the pixel as  a saturated pixel.
The Lyman-$\alpha$ optical depth, $\tau_{\rm alpha}$ in the considered
pixel at $\lambda^{\rm obs}_{\rm alpha}$,  is taken to be the smallest
of the above $\tau_{\rm series}$ that are larger than the noise rms at
the corresponding  position in the spectrum. This  procedure allows us
(i) to estimate  the optical depth even if  a transition is saturated;
(ii) to avoid part of  the blending effects.  When all transitions are
saturated, the pixel is flagged and  is given a lower limit on $\tau$.
Once $\tau_{\rm  alpha}$ is estimated over the  above wavelength range
[$\lambda_{\rm min}$,$\lambda_{\rm max}$],  it is possible to subtract
to  the  spectrum  the  corresponding  optical  depths  in  all  Lyman
transitions.  Then it  is possible to go on  with the determination of
$\tau_{\rm   alpha}$   over   the  wavelength   range   [$\lambda_{\rm
min}$,$\lambda_{\rm                max}$]~=~               [(1+$z_{\rm
em}$)$\times$1025.67$^2$/1215.67,   (1+$z_{\rm   em}$)$\times$1025.67]
using the same procedure.
\par\noindent
The advantage of  this procedure is not only  to increase the redshift
range  over  which  the  study  can  be performed  but  to  clean  the
Lyman-$\alpha$  forest  from  most   of  the  polluting  strong  metal
absorptions. Indeed,  the strong metal absorptions are  left over once
the H~{\sc  i} absorptions have  been subtracted.  Figure~\ref{filter}
illustrates  how  successful  the  procedure  can  be  in  recognising
automatically strong  absorptions due to  metals in the forest  on the
basis of the forest properties only.
\subsection{Cleaning the spectra}
The spectra have been scrutinised to flag all portions of the lines of
sight where  strong metal  lines hide the  information we  need. These
regions  and their complements  (O~{\sc vi}  and H~{\sc  i} wavelength
ranges  in the forest,  C~{\sc iv}  wavelength range  in the  red) are
removed from the analysis.  This includes in particular strong Fe~{\sc
ii} and Mg~{\sc ii} systems and all associated systems. The latter are
characterised  by  partial   covering  factors  and  large  $N$(C~{\sc
iv})/$N$(H~{\sc  i}),  $N$(N~{\sc  v})/$N$(H~{\sc i})  and  $N$(O~{\sc
vi})/$N$(H~{\sc  i}) column density  ratios.  When  in doubt,  we have
kept the absorptions  as in the system shown  on figure ~\ref{he0151}.
The   latter  is   observed  at   $z_{\rm  abs}   =   2.4178$  towards
HE0151$-$4326, 42600~km.s$^{-1}$  from the emission  redshift ($z_{\rm
em}$ = 2.789).

This means that the C~{\sc iv}  and O~{\sc vi} optical depths could be
slightly overestimated although we  are confident that enough care has
been exercised  and no obvious system  has been missed.   We have also
restricted   the   analysis    to   absorptions   located   at   least
3000~km~s$^{-1}$ away from the emission redshift.
\begin{figure}
\centering
\resizebox{0.9\hsize}{!}{\includegraphics{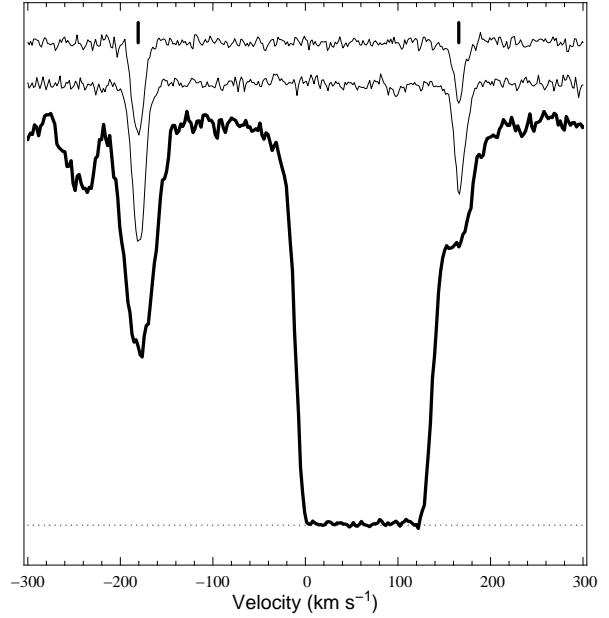}}
\caption{Portion  of  the   spectrum  of  HE~0151$-$4326  centered  at
  $z$=2.4178  showing a  strong Lyman-$\alpha$  absorption line  with no
  C~{\sc  iv}  absorption  associated  but two  satellites  H~{\sc  i}
  absorptions  have  strong  associated  C~{\sc  iv}  and  O~{\sc  vi}
  absorptions.   These satellites  are  narrow, there  is  no sign  of
  strong displacement  between H~{\sc i},  C~{\sc iv} and  O~{\sc vi},
  N~{\sc v}  is absent and covering  factor is unity:  these lines are
  kept in the analysis and are not considered as "associated" with the
  quasar.}
\label{he0151}
\end{figure}
\subsection{Metal optical depth}
Once the optical depth in the Lyman-$\alpha$ forest, $\tau_{\rm HI}$, 
is known, the corresponding optical depth of associated metal 
transitions at the same redshift can be derived. 
\par\noindent
To  avoid the  effects of  blending, we  use doublets  such  as C~{\sc
iv}$\lambda\lambda$1548,1550  and  O~{\sc vi}$\lambda\lambda$1031,1037
to secure the optical depth. For  each of the doublets, the first line
(\#1  of rest  wavelength  $\lambda^0_1$) has  an oscillator  strength
twice   larger  than  the   second  line   (\#2  of   rest  wavelength
$\lambda^0_2$). Therefore,  the true  optical depth in  transition \#1
must be twice larger than the true optical depth in transition \#2. We
therefore can impose that the  two optical depths be consistent. Using
the observed  optical depths at  the position of the  two transitions,
$\tau^{\rm obs}_1$  and $\tau^{\rm obs}_2$,  we want to  determine the
best  estimator   of  the  optical  depth,  $\tau_1$,   in  the  first
transition. Starting  from the  smallest wavelength, we  calculate for
each pixel  of wavelength $\lambda_1$,  the observed optical  depth at
this  position  $\tau^{\rm  obs}_1$~=~$-$ln($F_1$)  and  the  observed
optical   depth  $\tau^{\rm   obs}_2$~=~$-$ln($F_2$)  at   a  position
$\lambda_2$~=~$\lambda_1\times \lambda^0_2$/$\lambda^0_1$.
\par\noindent
Two cases  require special consideration. The  potential absorption is
said to  be "{\sl  In the noise}"  if $F >  1-\kappa\times\sigma$ with
$\sigma$  the  rms  of  the  noise  in  the  corresponding  pixel  and
$\kappa$~=~3.  When  $F < \kappa\times\sigma$, the  absorption is said
to be "{\sl Saturated}".
\par\noindent
We  use   for  $\tau_1$,  either  $\tau^{\rm   obs}_1$  or  $\tau^{\rm
obs}_2\times f_1/f_2$ depending on the status of the absorptions.  The
different cases  are summarised  in Table~\ref{taumetal}. In  the case
the two  lines are neither in  the noise nor saturated  (the lines are
"{\sl OK}"), or if  \#1 is "{\sl OK}" and \#2 is  "{\sl In the noise}"
(cases denoted "{\sl  Check}" in the Table), some  additional test has
to be performed.  First $F_1$  is used to predict $F^{\rm exp}_2$ that
is the  expected flux $F_2$  in case the  absorptions are only  due to
unblended lines  from the doublet  considered. If $F^{\rm  exp}_2 \geq
(F_2-\sigma_2)$,   then   $\tau_1$~=~$\tau^{\rm   obs}_1$.   If   not,
$\tau_1$~=~$\tau^{\rm obs}_2\times f_1/f_2$.
Once  $\tau_1$  is defined  and  if  $F_{\rm 1}$~$<$~$1-\kappa  \times
\sigma_1$, we  replace $\tau_2^{\rm obs}$ at  position $\lambda_2$, by
$\tau_2^{\rm  obs} -  \tau_1\times  f_2/f_1$.  This  corresponds to  a
self-contamination correction.
\begin{table}
\caption {Determination of $\tau_1=\tau_{\rm metal}$}
\begin{tabular}{ccccc}
\hline
  &   & \multicolumn{3}{c}{Line 1}   \\
  &                   & Saturated   & In the noise & OK  \\
\hline
  &         Saturated & Lower limit &  $\tau_1^{\rm obs}$    & $\tau_1^{\rm obs}$ \\
Line 2 & In the Noise & $\tau_2^{\rm obs}\times f_1/f_2$ & $\tau_1^{\rm obs}$ & Check \\
& OK & $\tau_2^{\rm obs}\times f_1/f_2$ & $\tau_1^{\rm obs}$ & Check \\
\hline
\end{tabular}
\label{taumetal}
\end{table}
\par\noindent
\subsection{Simulations}
To  check  our  two  procedures,  normalisation of  the  spectrum  and
recovery of  the C~{\sc  iv} optical depth,  we simulated  QSO spectra
drawing H~{\sc  i} absorption lines  at random from a  population with
the  same  column  density  and  Doppler  parameter  distributions  as
observed, using the  same spectral resolution as in  the UVES data and
adding noise. We use S/N~=~10000 to simulate perfect data and S/N~=~70
to simulate observations. We include C~{\sc iv} absorption at the same
redshift   as   H~{\sc   i},   assuming  a   fixed   log~$\tau$(C~{\sc
iv})/$\tau$(H~{\sc  i}) ratio.  We  recover $\tau$(C~{\sc  iv}) versus
$\tau$(H~{\sc   i})  and  plot   results  in   Fig.~\ref{simuciv}  for
log~$\tau$(C~{\sc  iv})/$\tau$(H~{\sc   i})~=~$-$2.5  (squares),  $-$3
(crosses),  $-$3.5  (circles) and  $-$10  (diamond)  to simulate  data
without any enrichment.  We  applied the same pixel-to-pixel procedure
after normalisation of  the data either with the  known continuum (top
panels) or using our normalisation procedure (bottom panels).  We plot
$\tau$(C~{\sc  iv}) recovered  by the  procedure  versus $\tau$(H~{\sc
i}). It can  be seen that we can  recover correctly $\tau$(C~{\sc iv})
in all  cases down to  log~$\tau_{\rm HI}$~$\sim$~0.  The  presence of
metals  is  revealed by  a  correlation  between  $\tau_{\rm HI}$  and
$\tau_{\rm metals}$ and the plateau  at small H~{\sc i} optical depths
is  a  consequence of  limited  S/N ratio.   It  is  clear that  these
simulations give us full confidence in our results.
\begin{figure}
  \centering
  \resizebox{0.95\hsize}{!}{\includegraphics[angle=-90]{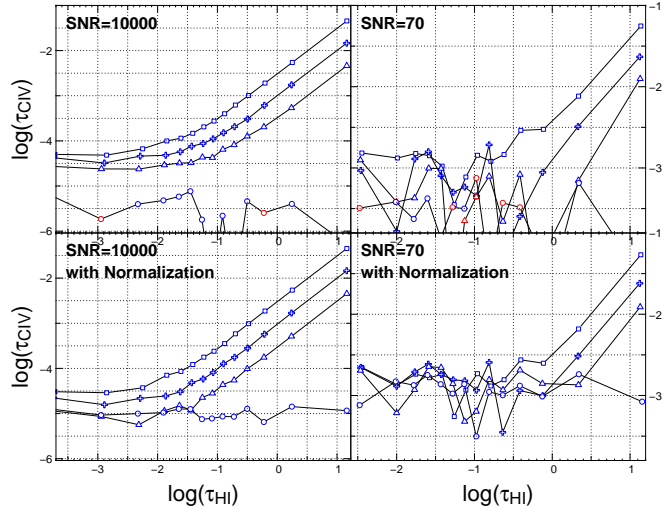}}
  \caption{Test  of the  normalisation and  pixel-by-pixel procedures.
We simulate QSO  spectra drawing H~{\sc i} absorption  lines at random
from a population  with the same column density  and Doppler parameter
distributions as  observed, using the same spectral  resolution as the
UVES  data  and adding  noise,  S/N~=~10000  in  the left  panels  and
S/N~=~70 in the right panels.  We include C~{\sc iv} absorption at the
same   redshift   as   H~{\sc    i}   assuming   some   fixed   ratio,
log~$\tau$(C~{\sc  iv})/$\tau$(H~{\sc   i})~=~$-$2.5  (squares),  $-$3
(crosses), $-$3.5 (triangles) and $-$10 (diamonds).  We apply the same
pixel-to-pixel   procedure   (all   panels)   to  the   data   without
normalisation (the continuum is assumed to be known; upper panels) and
with  normalisation (using  our  automatic procedure)  of the  spectra
(bottom panels). We plot $\tau$(C~{\sc iv}) recovered by the procedure
versus $\tau$(H~{\sc i}). It can be seen that we can recover correctly
$\tau$(C~{\sc iv}) in all cases down to log~$\tau$~$\sim$~0.  }
\label{simuciv}
\end{figure}
\section{Results}
\begin{figure*}
\centering
\resizebox{12cm}{!}{\includegraphics[angle=-90]{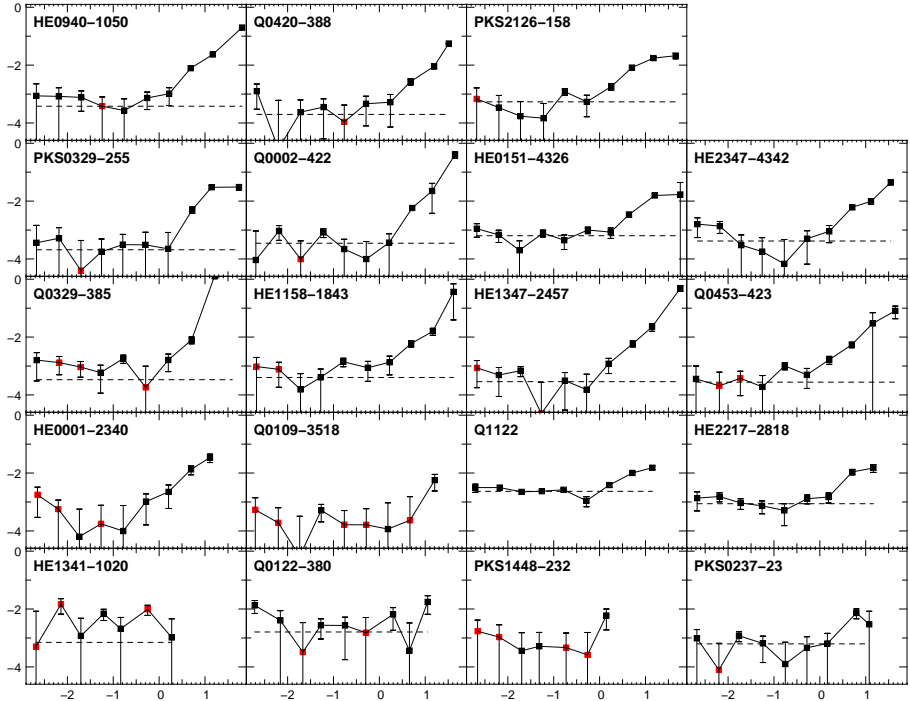}}
\caption{C~{\sc iv} optical depth versus H~{\sc i} optical depth
for individual quasars in the sample. 
}
\label{civind}
\end{figure*}
\begin{figure*}
\centering
\resizebox{12cm}{!}{\includegraphics[angle=-90]{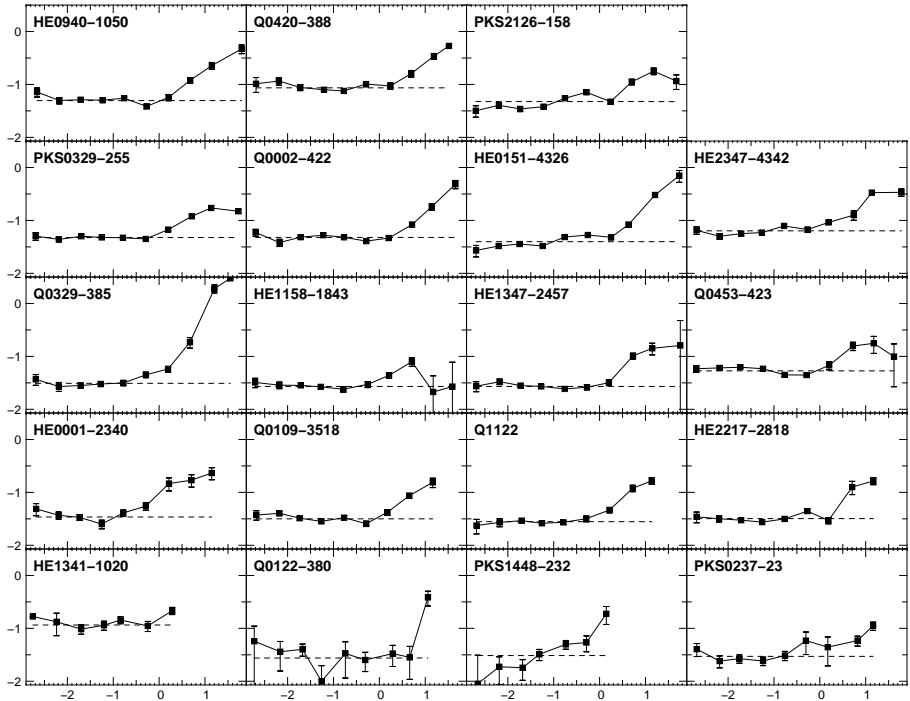}}
\caption{O~{\sc vi} optical depth versus H~{\sc i} optical depth
for the 19 individual quasars in the sample.
}
\label{oviind}
\end{figure*}
We apply the  following procedure either for one line  of sight or for
the complete  data set, combining all  lines of sight.   Pixels in the
Lyman-$\alpha$  forest are  sorted in  order of  increasing  H~{\sc i}
Lyman-$\alpha$ optical depth and  are gathered in predefined bins.  In
each of the bins, the  median of the corresponding $\tau_{\rm metal}$,
measured at the  same redshift, is then calculated.   Using the median
instead  of  the mean  avoids  the measurement  to  be  biased by  the
presence  of  a  few  strong  absorptions. Individual  plots  for  the
nineteen lines  of sight are  shown for C~{\sc  iv} and O~{\sc  vi} in
Fig.~\ref{civind} and \ref{oviind}, respectively.
\subsection{Presence of C~{\sc iv}}
\begin{figure}
\centering
\resizebox{0.9\hsize}{!}{\includegraphics{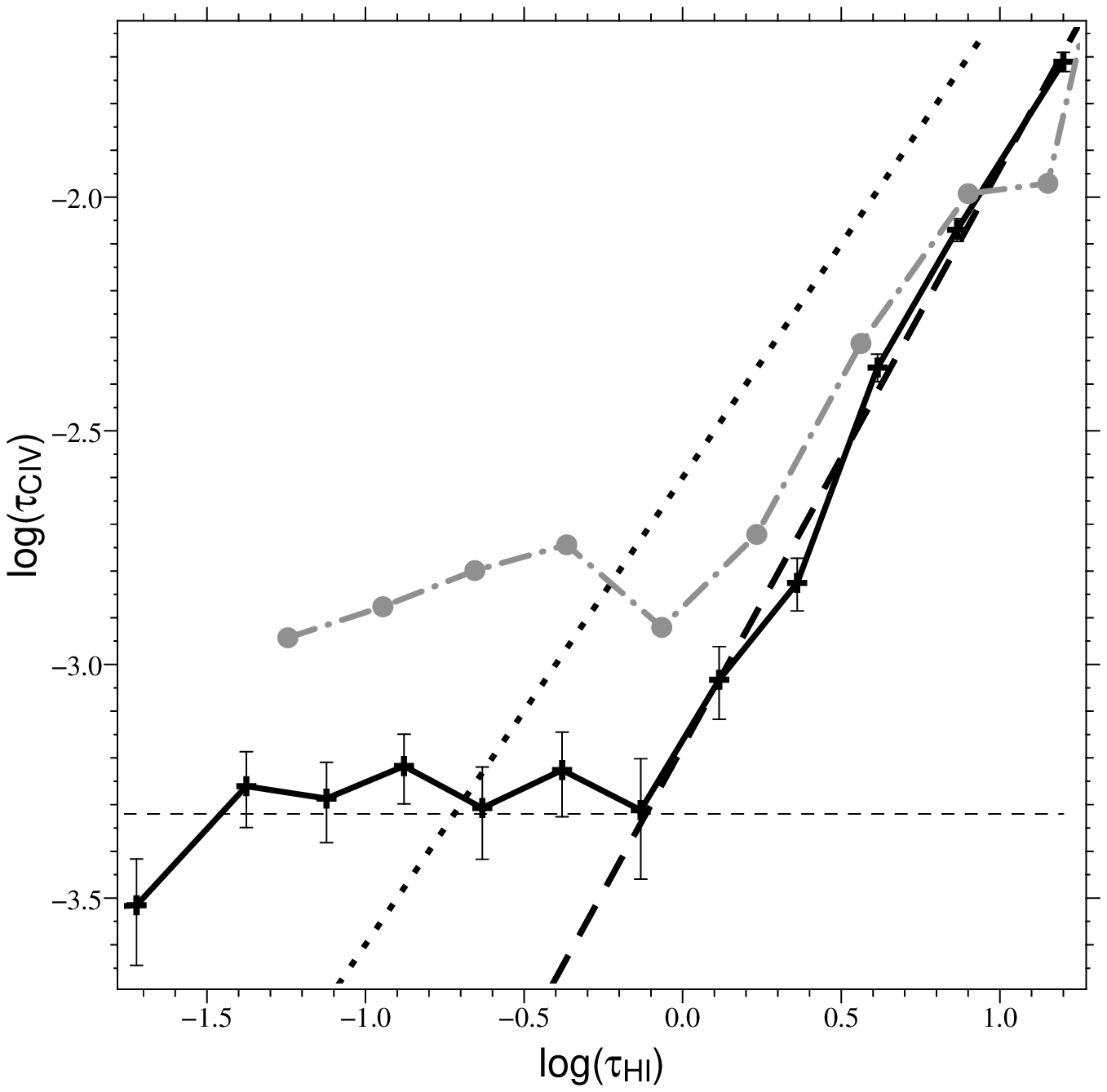}}
\caption{Median of C~{\sc iv} optical depth versus median of H~{\sc i}
optical depth is plotted with  black cross.  Only pixels with velocity
difference  larger  than  3000~km~s$^{-1}$  compared to  the  emission
redshift  of the  quasar  are  used. Grey  circles  correspond to  the
previous finding by Ellison et  al.  (2000).  The dotted line is drawn
for   illustration   and    corresponds   to   log~C~{\sc   iv}/H~{\sc
i}~$\sim$~$-$2.6 .   The dashed thick  line is a  fit to the  data for
$\log       \tau_{\rm      HI}       >       0$;      $\log\,\tau_{\rm
CIV}\,=\,1.3\times\log\,\tau_{\rm HI}-3.2$. The horizontal dashed line
corresponds  to  the median  of  $\tau_{\rm  CIV}$ for  log~$\tau_{\rm
HI}$~$<$~$-$1.3.}
\label{civtot}
\end{figure}
We  plot in  Fig.~\ref{civtot} the  median of  the C~{\sc  iv} optical
depth versus the median of  the H~{\sc i} optical depth when combining
the whole sample.
It can  be seen that there is  an excess of C~{\sc  iv} for $\tau_{\rm
HI}$~$>$~1.   This excess  is present  for nearly  all  the individual
lines  of sight (see  Fig.~\ref{civind}).  For  $\tau_{\rm HI}$~$<$~1,
there  is no  detection of  C~{\sc iv}.   In particular,  the apparent
optical depth is  statistically the same and consistent  with zero for
all the bins corresponding to log~$\tau_{\rm HI}$~$<$~0.
\par\noindent
The  thick  dashed  line  corresponds   to  a  fit  to  the  data  for
log~$\tau_{\rm                HI}$~$>$~0:               log~$\tau_{\rm
CIV}$~=~1.3$\times$log~$\tau_{\rm    HI}$~$-$~3.2.     The   dotted
diagonale  line   is  drawn   for  illustration  and   corresponds  to
log($\tau_{\rm CIV}/\tau_{\rm HI}$)~=~$-$2.6. It  can be seen that the
diagonale line is  rejected by our data. Indeed,  the present data are
inconsistent  with  a  uniform  distribution  of  C~{\sc  iv}  in  the
Lyman-$\alpha$  forest with  constant  $\tau_{\rm CIV}/\tau_{\rm  HI}$
ratio.  On the  contrary our results strongly suggest  that the C~{\sc
iv}/H~{\sc i}  ratio increases with increasing $\tau_{\rm  HI}$. It is
also  apparent  that   for  log~$\tau_{\rm  HI}$~$\sim$~0,  log(C~{\sc
iv}/H~{\sc  i})~$<$~$-$3.2.    Although  the  results   by  Schaye  et
al. (2003) favor a slope closer to the diagonale than what we observe,
their error bars are larger and their measurements are consistent with
our findings.
\par\noindent
It  is important  to note  also  that the  scatter in  the C~{\sc  iv}
optical depth is  very similar at different H~{\sc  i} optical depths.
This means that  although the mean C~{\sc iv}  optical depth decreases
with decreasing  $\tau_{\rm HI}$, large C~{\sc iv}  optical depths can
be seen at nearly any H~{\sc  i} optical depth. This is illustrated in
Fig.~\ref{he0151} where a strong Lyman-$\alpha$ system is seen with no
C~{\sc  iv}   absorption  associated  down  to   our  detection  limit
$\sim$10$^{12}$~cm$^{-2}$,  but two  satellite  H~{\sc i}  absorptions
have strong associated C~{\sc iv}  and O~{\sc vi} absorption (see also
Bergeron et al. 2002). It must  be noted that these systems have quiet
dynamics  (the different  absorptions of  C~{\sc iv},  O~{\sc  vi} and
H~{\sc i} are fairly well centered on top of each others), they do not
show evidence for partial covering  factors and have very little or no
N~{\sc v} associated  absorption. Therefore, their properties strongly
differ from those of systems  associated with the quasar and they must
be truly intervening systems.
\subsection{Presence of OVI}
\begin{figure}
\centering
\resizebox{0.9\hsize}{!}{\includegraphics{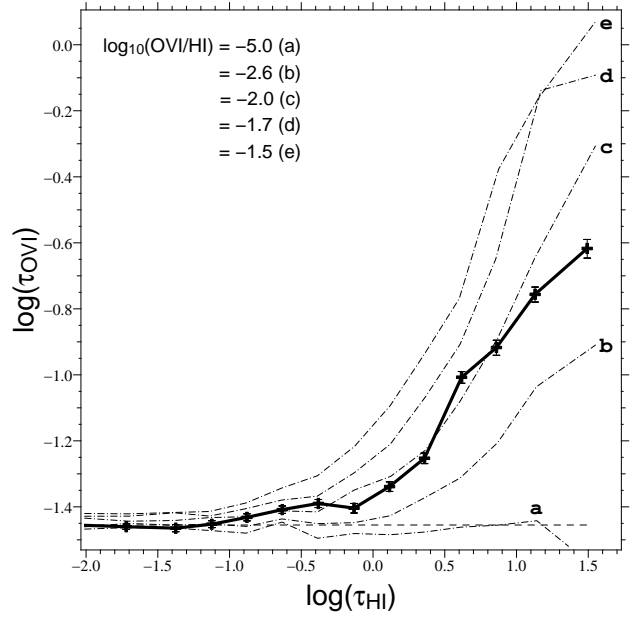}}
\caption{The  median O~{\sc vi}  optical depth  is plotted  versus the
median H~{\sc  i} optical  depth. The crosses  are the  observed data,
other curves are simulations  where the ratio log~O~{\sc vi}/H~{\sc i}
is  given fixed values,  $-$5.0 (no  metals), $-$2.5,  $-$2.0, $-$1.7,
$-$1.5  from  the  curve  labelled   (a)  to  the  one  labelled  (e),
respectively}
\label{ovitot}
\end{figure}
Schaye et al.  (2000) performed  a similar pixel-by-pixel search for O
VI  absorption  in  eight  high-quality quasar  spectra  spanning  the
redshift range $z$ = 2.0--4.5.  In the redshift range 2~$<$~$z$~$<$~3,
they stated to detect O~{\sc vi} in the form of a positive correlation
between  the H~{\sc i}  Lyman-$\alpha$ optical  depth and  the optical
depth  in  the corresponding  O~{\sc  vi}  pixel,  down to  $\tau_{\rm
HI}$~$\sim$~10$^{-1}$.  On  the contrary,  they did not  detect O~{\sc
vi} at  $z >$~3  and considered this  is consistent with  the enhanced
photoionization from a  hardening of the UV background  below $z < 3$,
although this could also be  caused by the high level of contamination
from Lyman series lines.
\begin{figure}
\centering
\resizebox{0.9\hsize}{!}{\includegraphics{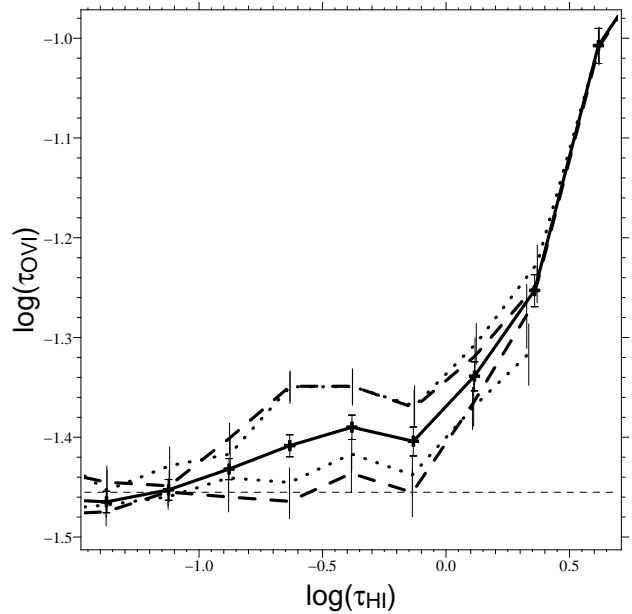}}
\caption{Median O~{\sc  vi} optical depth versus median  H~{\sc i} for
pixels located  at less than  500~km~s$^{-1}$ (upper dashed  curve) or
less than 300~km~s$^{-1}$ (upper dotted curve) from strong absorptions
($\tau_{\rm   HI}~>~4$)  and   for   pixels  located   at  more   than
500~km~s$^{-1}$  (lower  dashed curve)  or  more than  300~km~s$^{-1}$
(lower dotted curve) from strong absorptions.}
\label{oviv}
\end{figure}
\par\noindent
\par\noindent
It  can be  seen  from  Fig.~\ref{ovitot} that  our  data confirm  the
detection of O~{\sc vi} down  to $\tau_{\rm HI}$~$\sim$~0.2 that is at
smaller H~{\sc i} optical depth than for C~{\sc iv} which is seen down
to $\tau_{\rm  HI}$~$\sim$~1 only.  To  investigate if the  signal for
0.2~$<$~$\tau_{\rm  HI}$~$<$~1 comes  from all  parts of  the spectrum
however, we  have divided the  sample into two subsamples.   The first
subsample contains all H~{\sc i} pixels located within $\Delta v$ from
a  strong  absorption  with   $\tau_{\rm  HI}$~$>$~4  and  the  second
subsample contains all other pixels.  We then vary $\Delta v$.  It can
be seen on Fig.~\ref{oviv} that for $\Delta v$~$\leq$~300~km~s$^{-1}$,
the signal significantly increases for the first subsample. This means
that  O~{\sc vi}  is  predominantly  seen in  the  vicinity of  strong
Lyman-$\alpha$  absorption lines.  Indeed,  the signal  disappears for
the  second   subsample;  there  is  no  O~{\sc   vi}  absorption  for
0.2~$<$~$\tau_{\rm HI}$~$<$~1.  This new result strongly suggests that
the O~{\sc vi}  absorption arises mostly in regions  close in space to
those responsible for  strong H~{\sc i} absorptions and  a most likely
explanation is that the O~{\sc vi} phase is part of winds flowing away
from overdense regions.
\par\noindent
To derive a quantitative limit on  the amount of O~{\sc vi} present in
the IGM, we have performed  simulations of the metal enrichment of the
Lyman-$\alpha$  forest.    Artificial  spectra  are   created  drawing
absorption lines at  random from a population with  column density and
Doppler  parameter   distributions  consistent  with   those  observed
(Petitjean et al.  1993, Hu et  al. 1995, Kirkman and Tytler 1997, Kim
et  al. 2000).  The  number of  lines  is adjusted  so  that the  mean
absorption of the  simulated spectra is the same  as the observed one.
Although  this  is  probably   a  rough  assumption,  constant  O~{\sc
vi}/H~{\sc  i} ratio  is assumed  and O~{\sc  vi} absorption  is added
accordingly. Noise  consistent with  the data is  added and  the whole
procedure described in Section~3  is applied to the simulated spectra.
Results are  plotted in Fig.~\ref{ovitot}.   It can be seen  that data
are  consistent  with log~O~{\sc  vi}/H~{\sc  i}~$\sim$~$-$2, a  value
within the range found for  individual O~{\sc vi} systems (Bergeron et
al.  2002).
\begin{figure}
\resizebox{0.9\hsize}{!}{\includegraphics{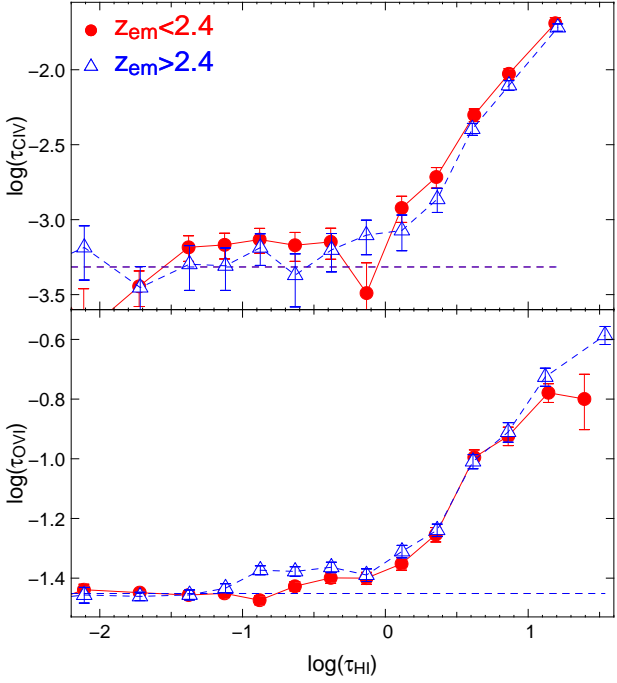}}%was width=8.cm
\caption{The  C~{\sc iv}  (top panel)  and O~{\sc  vi}  (bottom panel)
optical  depth is  plotted  versus  the H~{\sc  i}  optical depth  for
$z$~$<$~2.4 (filled circles and solid line) and $z$~$>$~2.4 (triangles
and dashed line).  }
\label{zevol}
\end{figure}
\par\noindent
\subsection{Evolution with redshift}
\begin{figure}
\centering
\resizebox{0.9\hsize}{!}{\includegraphics{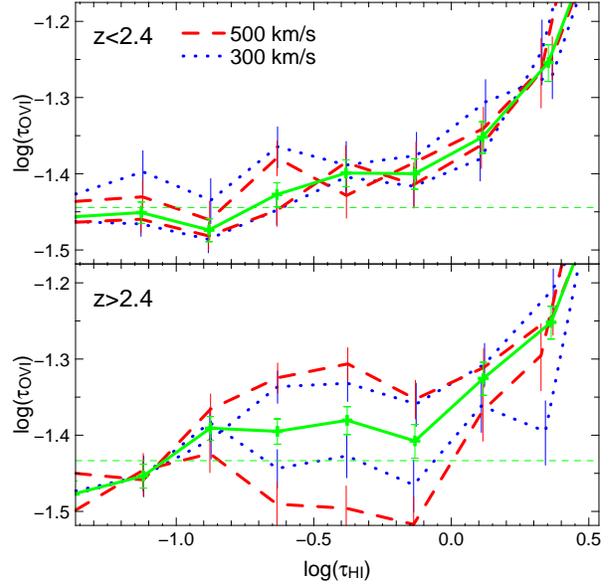}}
\caption{ Same as figure \ref{oviv} for the two redshift bins, $z$~$<$
and $>$~2.4.}
\label{oviz}
\end{figure}
We have divided the sample in two redshift bins with similar number of
pixels in each  bin, $z$~$>$ and $<$~2.4. We  plot in Fig.~\ref{zevol}
C~{\sc iv}  (top panel) and  O~{\sc vi} (bottom panel)  optical depths
versus  H~{\sc i} optical  depth for  $z$~$<$~2.4 (filled  circles and
solid line)  and $z$~$>$~2.4 (triangles  and dashed line).  It  can be
seen that C~{\sc iv} optical depth shows mild cosmic evolution when no
evolution  is  detected in  $\tau_{\rm  OVI}$.   The most  interesting
feature  however is that  the excess  of O~{\sc  vi} optical  depth at
log~$\tau_{\rm  HI}$~$\sim$~$-$0.5  is  more  important  in  the  high
redshift bin. Note that this signal is due to the three lines of sight
toward quasars  with the highest emission redshifts.   Moreover, if we
apply the same procedure as  previously, we find that, at $z$~$<$~2.4,
the O~{\sc vi} absorption does not  seem to be related to strong lines
when the relation is apparent at larger redshift (see Fig~\ref{oviz}).
As this signal  is probably related to galactic winds,  this may be an
indication that galactic winds are more prominent at higher redshift.
\section{Conclusion}
We have used a pixel-by-pixel  analysis to investigate the presence of
metals  in the  inter galactic  medium.  We  have measured  the median
optical depth  of C~{\sc iv}$\lambda$1548  and O~{\sc vi}$\lambda$1031
absorptions versus  the H~{\sc i} Lyman-$\alpha$ optical  depth of the
gas at a  mean redshift $z$~=~2.6 using 19 lines  of sight observed at
high spectral  resolution ($R$~=~45000)  and high and  homogeneous S/N
ratio  (approximately 35  and  70 per  pixel  over, respectively,  the
O~{\sc vi}  and C~{\sc  iv} wavelength ranges).   Great care  has been
exercized to determine the continuum.
\par\noindent
We have  carefully determined  the H~{\sc i}  optical depth  by taking
into   account  the   information   in  the   Lyman  series   (usually
Lyman-$\beta$ and Lyman-$\gamma$). We have also carefully removed from
the  spectra  all  wavelength  ranges  that  are  polluted  by  strong
intervening absorptions from metal line systems, in particular Mg~{\sc
ii} and Fe~{\sc ii} systems, and from associated systems.
\par\noindent
We find that  the gas is enriched in carbon  and oxygen for $\tau_{\rm
HI}$~$>$~1.  Contrary to previous  claims, there is no indication that
C~{\sc iv}  absorption is statistically associated with  gas of H~{\sc
i}  Lyman-$\alpha$ optical  depth smaller  than 1.   In  addition, our
observations  strongly suggest  that  the C~{\sc  iv}/H~{\sc i}  ratio
decreases  with  decreasing  $\tau_{\rm  HI}$.  We  observe  that  for
$\tau_{\rm      HI}$~$\sim$~1,     log~$\tau_{\rm     CIV}$/$\tau_{\rm
HI}$~$<$~$-$3.2    which   corresponds   to    log~C~{\sc   iv}/H~{\sc
i}$<$~$-$3.3 assuming  $b_{\rm HI}/b_{\rm CIV}=2$ and  a Voigt profile
or  log~C~{\sc  iv}/H~{\sc  i}$<$~$-$3.0 assuming  $b_{\rm  HI}=b_{\rm
CIV}$.  This does  not prevent a small fraction  of the low $\tau_{\rm
HI}$ gas  from being associated with  strong metal lines  (see Fig. 4)
indicating that enrichment is highly inhomogeneous.
\par\noindent
We  detect the  presence of  O~{\sc vi}  for  $\tau_{\rm HI}$~$>$~0.2,
consitent  with  a   constant  ratio  log~$\tau_{\rm  OVI}$/$\tau_{\rm
HI}$~$\sim$~$-$2.0    corresponding     to    log~O~{\sc    vi}/H~{\sc
i}~$\sim$~$-1.75$  assuming  $b_{\rm HI}/b_{\rm  OVI}=2$  and a  Voigt
profile or log~O~{\sc vi}/H~{\sc i}~$\sim$~$-1.45$ assuming $b_{\rm
HI}=b_{\rm OVI}$ . We show that for 0.2~$<$~$\tau_{\rm HI}$~$<$~1, the
O~{\sc  vi}   absorption  is   associated  with  gas   located  within
$\sim$300~km~s$^{-1}$ from strong H~{\sc i} lines.  This suggests that
the  O~{\sc vi}  phase is  probably part  of winds  flowing  away from
overdense regions.
\par\noindent
It is  therefore not possible to  conclude that metals  are present in
the   most  teneous   regions  of   the  IGM   ($\tau_{\rm  HI}$~$<$~1
corresponding to  overdensities of about 1  to 3 at  $z$~$\sim$3 and 2
respectively) far away from overdense regions.
%

%
%%%%%%%%%%%%%%%%%%%%%%%%%%%%%%%%%%%%%%%%%%%%%%%%%%%%%%%%%%%%%%%%%%%
%\vskip 0.5cm
%\noindent{\sl  Acknowledgements:} {
\acknowledgements{
We are grateful to the ESO support astronomers who have performed
the observations in service mode. We thank F. Primas for her help with the OBs.
We thank Evan Scannapieco and R. 
Srianand for useful comments on the manuscript.
}

\end{document}